# SimiNet: a Novel Method for Quantifying Brain Network Similarity

Ahmad Mheich[1, 2, 3], Mahmoud Hassan[1, 2], Mohamad Khalil[3], Vincent Gripon[4], Olivier Dufor[4] and Fabrice Wendling[1, 2]

[1] INSERM, U1099, Rennes, F-35000, France

[2] Université de Rennes 1, LTSI, F-35000, France

[3] AZM center-EDST, Lebanese University, Tripoli, Lebanon

[4] Telecom Bretagne (Institut Mines-Telecom), UMR CNRS Lab-STICC, Brest

Corresponding author:

Mahmoud Hassan

mahmoud.hassan@univ-rennes1.fr

Tel: +33 2 23 23 56 05 / 62 20

Fax: +33 2 23 23 69 17

# Abstract

Quantifying the similarity between two networks is critical in many applications. A number of algorithms have been proposed to compute graph similarity, mainly based on the properties of nodes and edges. Interestingly, most of these algorithms ignore the physical location of the nodes, which is a key factor in the context of brain networks involving spatially defined functional areas. In this paper, we present a novel algorithm called "SimiNet" for measuring similarity between two graphs whose nodes are defined a priori within a 3D coordinate system. SimiNet provides a quantified index (ranging from 0 to 1) that accounts for node, edge and spatiality features. Complex graphs were simulated to evaluate the performance of SimiNet that is compared with eight state-of-art methods. Results show that SimiNet is able to detect weak spatial variations in compared graphs in addition to computing similarity using both nodes and edges. SimiNet was also applied to real brain networks obtained during a visual recognition task. The algorithm shows high performance to detect spatial variation of brain networks obtained during a naming task of two categories of visual stimuli: animals and tools. A perspective to this work is a better understanding of object categorization in the human brain.

# Introduction

The brain is a large-scale network in which distant interconnected neural assemblies continuously synchronize and desynchronize to process information. Over the past decade, considerable progress has been achieved in neuroimaging techniques that are now able to identify brain networks from structural and/or functional data. In most cases, these networks are represented as graphs in which the nodes denote brain areas and the edges describe the connectivity among these areas (Bullmore and Sporns, 2009). This graph representation allows for application of graph theory algorithms in order to assess statistical and/or topological properties of networks reconstructed from data. From a theoretical viewpoint, the application of these algorithms on functional, as well as on structural connectivity matrices, has revealed many properties about brain networks, such as small-worldness (Achard et al., 2006; Bassett and Bullmore, 2006), modularity (Bassett et al., 2011; Meunier et al., 2010), hubs (Hagmann et al., 2008), and rich-club configurations (van den Heuvel and Sporns, 2011). From an application viewpoint, graph theory based analysis has been widely used to characterize normal (Bressler and Menon, 2010) and pathological (Fornito et al., 2015) brain activities from several modalities (fMRI, EEG, MEG). In particular, it has led to the identification of network alterations in aging (Gong et al., 2009), Alzheimer's disease (Stam et al., 2007), schizophrenia (Liu et al., 2008) and autism(Guye et al., 2010).

In contrast with the large number of methods aiming at characterizing graph properties (Costa et al., 2007), less attention has been paid to methods able to quantitatively compare graphs and extract similarities while taking into account the spatial location of the graph nodes. Reported methods make use of graph kernels (Borgwardt and Kriegel, 2005; Shervashidze et al., 2011; Vishwanathan et al., 2010), graphs and subgraphs isomorphism (Cordella et al., 2004), graph edit distance (Gao et al., 2010) and Levenshtein distance (Cao et al., 2013) in order to measure graph similarity. However, in the context of brain networks, the spatial location of nodes is a key factor for the graph comparison (Pineda-Pardo et al., 2014). The intuitive idea is that two networks with strictly identical statistical properties but interconnecting different brain areas should be considered to have low similarity. Conversely, two graphs with partly dissimilar properties but interconnecting the same brain regions should be considered as exhibiting high similarity.

In this paper, we propose a new algorithm to solve this issue. This algorithm is able to measure the similarity between two graphs based on the node and edge properties but under a spatial constraint related to the physical location of nodes. To our knowledge, this approach has never been followed in the problem of measuring a formal distance between brain networks. The performance of the proposed algorithm was evaluated using simulated graphs as well as real brain networks estimated from dense-electroencephalographic (EEG) signals during a picture naming task involving two categories of stimuli: tools and animals. The paper is organized as follows. Section II.A provides the notations and definitions. The problem statement is then introduced in section II.B followed by the description, in section II.C, of the proposed method for measuring graphs similarity. A comparative study is then achieved with respect to already-published methods briefly presented in section II.D. Results obtained on simulated networks as well as on real networks are given in section III. Finally, these results are discussed according to the performance and potential applications of the proposed algorithm in the context of brain connectivity research.

# Materiels and Methods

## A. Notations and definitions

A graph $G$ is denoted by $G(V, E)$ where $V$ denotes the set of nodes (with known Cartesian coordinates) and $|V|$ is the order of the graph (number of nodes). $E \subseteq V \times V$ defines the edges and $|E|$ is the number of edges in $G$. We denote by $W_{v_1,v_2}^G$ the weight of the edge between nodes $v_1$ and $v_2$. The graph is said to be simple if there is no edge linking a node with itself, and it is said to be undirected if the adjacency matrix is symmetric. We denote by $sim(G_1, G_2)$ the similarity measure between two graphs $G_1$ and $G_2$. All graphs considered in this work are assumed to be simple, undirected, weighted graphs. Defined quantities and notations are listed in Table 1.

## B. Problem statement

Let us define two graphs $G_1(V_1, E_1)$ and $G_2(V_2, E_2)$. The nodes of $G_1$ and $G_2$ are distributed onto a square grid of side length $T$. Cells in the grid are indexed from 1 to $T^2$ line by line starting from the first line. We suppose that each grid cell contains at most one node with Cartesian coordinates $(x, y)$. Consequently, we index nodes using the corresponding grid cell index: $K_v = x_v + ((y_v - 1) \times T)$ where $x_v \in [1, T]$ and $y_v \in [1, T]$. The problem is two-fold: i) to elaborate a distance $d(G_1, G_2)$ that accounts for the positions of nodes respectively in $G_1$ and $G_2$ as well as for statistical properties (number of nodes, number and weights of edges, density). ii) To devise a normalized index $sim(G_1, G_2) \in [0,1]$ that globally reflects the similarity of compared graphs $G_1$ and $G_2$, the similarity index to be elaborated should take values in $[0,1]$.

Lowest values denote highly different graphs whereas values close to 1 denote almost identical graphs, in terms of statistical and spatial properties (position of nodes).

## C. Proposed method

In order to solve the above stated problem, we propose a new algorithm called SimiNet (described in Fig.1) which calculates a distance between $G_1$ and $G_2$ (Fig. 1A). This distance is based on i) the cost of a sequence of cost-based changes (substitution, insertion, deletion) on nodes that are necessary to map $G_1$ onto $G_2$ and ii) the computation of the difference of edge weights between $G_1$ and $G_2$.

The proposed algorithm includes four main steps, as illustrated in Fig. 1, B-E:

### Step 1: Node substitution

To start with, the nodes common to $G_1$ and $G_2$ that have the same spatial location are detected (these nodes have the same index value $K_v$ in $G_1$ and $G_2$). In the example of Fig. 1B-i, we observe three nodes of $G_1$ having the same spatial position as three nodes of $G_2$. For the remaining nodes, we define a spatial neighborhood $\Omega$ (disk with radius $R = 1.5$ as depicted in Fig. 1B-ii). The idea is to shift a neighbor node $v_1$ included in $V_1$ to the spatial position of $v_2$ included in $V_2$ if $v_1$ is located in $\Omega$ (the defined spatial neighborhood of $v_2$). In the case where there are $>1$ nodes of graph $G_1$ located in $\Omega$, the nearest node to $v_2$ is shifted. The cost of shifting is equal to the Euclidian distance between the two shifted nodes ($v_1$ and $v_2$). The node distance ($ND$) is updated by the Euclidian distance between $v_1$ and $v_2$ (e.g. $ND = 1$ in Fig. 1B-ii).

**Step 2: Node Insertion**

In step 2, the insertion operation is used to add a new node in $G_1$ at the same position of node $v_2$ in the case where no node exists in the defined spatial neighborhood $\Omega$ for each node $v_2$ of $V_2$. The cost of insertion must be higher than the cost of substitution. In the example of Fig. 1C, the cost of insertion was chosen equal to disk with radius $R = 1.5$. In this case, the node distance ($ND$) is updated to 1+1.5=2.5

**Step 3: Node deletion**

At the end of step 2, all the nodes that are included in $G_2$ are also included in $G_1$ (Fig. 1D). In step 3, the remaining nodes of $G_1$ are then deleted. The cost of deletion for one node is equal to the insertion cost. The node distance between $G_1$ and $G_2$ is updated accordingly, as exemplified in Fig. 1D where $ND = 2.5 + 3 = 5.5$ since two nodes are deleted. At this stage, $G_1$ and $G_2$ have the same number of nodes located at the same positions on the grid.

**Step 4: Edge distance**

In step 4, the edge distance ($ED$) for $G_1$ and $G_2$ at the initial state is computed from the difference of edge weights between $G_1$ and $G_2$, using (1):

$$diff(W^{G_1}_{v_1^p, v_1^k}, W^{G_2}_{v_2^p, v_2^k}) = \left| W^{G_1}_{v_1^p, v_1^k} - W^{G_2}_{v_2^p, v_2^k} \right| \qquad (1)$$

In a binary graph, the $diff(W^{G_1}_{v_1^p, v_1^k}, W^{G_2}_{v_2^p, v_2^k})$ score $\in [0,1]$ where 1 means that the edge linking $v_1^p$ and $v_1^k$ does exist in $G_1$ but does not exist in $G_2$ for the corresponding nodes $v_2^p$ and $v_2^k$. Conversely, a difference equal to 0 means that this edge exists in both graphs. The total $ED$ is then calculated using (2):

$$ED = \sum_{k=1}^{T-1} \sum_{p=k+1}^{T} diff(W^{G_1}_{v_1^p, v_1^k}, W^{G_2}_{v_2^p, v_2^k}) \qquad (2)$$

Finally, the distance between $G_1$ and $G_2$ is calculated by summing up the node distance ($ND$ from step 3) and the edge distance ($ED$ from step 4)

$$d(G_1, G_2) = ND + ED \qquad (3)$$

In the example of Fig. 1, the edge distance ($ED$) is equal to 9. The distance $d$ between $G_1$ and $G_2$ is calculated by the equation 3: $d(G_1,G_2) = ND + ED = 5.5+9=14.5$.

Finally, the distance $d(G_1,G_2)$ is scaled to a similarity index $sim_{SimiNet}(G_1,G_2)$ via the formula $sim_{SimiNet} = (1/(1+d))$. As depicted, the proposed similarity index $\in [0,1]$ where 0 means that $G_1$ and $G_2$ are totally dissimilar, while 1 means that $G_1$ and $G_2$ are identical.

The pseudo-code of this algorithm is provided in Fig. 2.

**D. Other reported algorithms**

In section 2.5, SimiNet is compared with eight state-of-art methods able to measure similarity between graphs. These eight methods are briefly described hereafter.

**Graph edit distance (GED)** (Bunke et al., 2007)

This distance is based on the transformation of one graph to the other using elementary operations. The elementary operations consist in suppressions and insertions of nodes.

$$d_{GED}(G_1,G_2) = |V_1| + |V_2| - 2|V_1 \cap V_2| + \\ |E_1| + |E_2| - 2|E_1 \cap E_2|$$

This distance is normalized between 1 and 0 via the formula $sim_{GED}(G_1,G_2) = (1/(1+d_{GED}(G_1,G_2)))$.

**DeltaCon method** (Koutra et al., 2013)

This algorithm assesses the similarity between two graphs on the same nodes. The concept of this method is to compute the pairwise node affinities in the first graph and to compare them with the ones in the second graph. Then, it measures the differences of nodes affinity scores of the two graphs and reports the similarity score. Readers may refer to (Koutra et al., 2013) for details about the DeltaCon algorithm main steps and implementation. This algorithm also provides a normalized similarity index $sim_{DeltaCon}(G_1,G_2)$ ranging from 0 (dissimilar graphs) to 1 (identical graphs).

**Vertex/Edge Overlap method (VEO)** (Papadimitriou et al., 2010).

The principle of this method is that two graphs are similar if they share many vertices (i.e. nodes) and edges. Thus, the similarity between two graphs $G_1(V_1, E_1)$ and $G_2(V_2, E_2)$ is defined as:

$$sim_{VEO}(G_1, G_2) = \frac{|V_1 \cap V_2| + |E_1 \cap E_2|}{|V_1| + |V_2| + |E_1| + |E_2|}$$

This measure of similarity is computed by scanning all nodes of $G_1$ and by checking if each occurs in $V_2$, the set of nodes of $G_2$.

**λ-distance method (Lambda distance)** (Wilson and Zhu, 2008)

Let $(\lambda_{1l})_{l=1}^{|V_1|}$ and $(\lambda_{2l})_{l=1}^{|V_2|}$ be the eigenvalues of two adjacency matrices respectively associated with two graphs $G_1$ and $G_2$. The λ-distance is given by:

$$d_\lambda(G_1, G_2) = \sqrt{\sum_{l=1}^{L} (\lambda_{1l} - \lambda_{2l})^2}$$

where $L$ is $\max(|V_1|, |V_2|)$. For the sake of comparison, this distance is also normalized between 1 and 0 via the formula $sim_\lambda(G_1, G_2) = (1/(1 + d_\lambda(G_1, G_2)))$

**Graph Kernel**

The kernel function is a popular method to find different types of relations between datasets. A graph kernel is the application of the kernel function to graphs where the objective is to find relationships (similarity) between two graphs. Several graph kernels algorithms have been proposed to measure network similarity such as random walks, shortest paths and Weisfeiler-Lehman. These methods are briefly described hereafter:

- *Random walk (RWkernel )*(Vishwanathan et al., 2010)**:** given two graphs $G_1$ and $G_2$, a random walk kernel counts the number of matching labeled random walks. The matching between two nodes is determined by comparing their attribute values. The measure of similarity between two random walks is then the product of the kernel values corresponding to the nodes encountered along the walk.

- *Shortest path kernel (spKernel)*(Borgwardt and Kriegel, 2005)**:** compute the shortest path kernel for a set of graphs by exact matching of shortest path lengths. The Floyd-Warshall algorithm (Floyd, 1962) is usually used to calculate all the pairs-shortest-paths in $G_1$ and $G_2$. The shortest path kernel is then defined by comparing all the pairs of the shortest path lengths among nodes in $G_1$ and $G_2$.

- *Weisfeiler-Lehman(WL)*(Shervashidze et al., 2011)**:** compute $h$-step Weisfeiler-Lehman kernel for a set of graphs. The main idea of this algorithm is to increase the node labels by the sorted set of node labels of neighboring nodes, and compress these increased labels into new shorted labels. These steps are repeated until the node label sets of $G_1$ and $G_2$ differ or the number of iterations reaches $h$.

- *Weisfeiler-Lehman shortest path kernel (WLspdelta)*(Shervashidze et al., 2011)**:** compute the $h$-step Weisfeiler-Lehman shortest path delta kernel between the compared graphs $G_1$ and $G_2$.

**E. Comparative analysis**

In order to compare the proposed algorithm with the state-of-art methods, we analyzed the performance of the nine methods using graphs subject to three types of alterations, namely changes in the edge weights, insertion of nodes and shifts in their spatial location. In practice, a random graph $G_1$ with 20 nodes located onto a grid $Gr(20 \times 20)$ was generated. This graph was altered to get a graph $G_2$ and the five similarity indexes ($sim_{SimiNet}(G_1,G_2)$, $sim_{GED}(G_1,G_2)$, $sim_{DeltaCon}(G_1,G_2)$, $sim_{VEO}(G_1,G_2)$, $sim_{\lambda}(G_1,G_2)$, $sim_{RW\,kernel}(G_1,G_2)$, $sim_{WL}(G_1,G_2)$, $sim_{sp\,kernel}(G_1,G_2)$ and $sim_{WLspdelta}(G_1,G_2)$) between the initial graph $G_1$ and the altered version $G_2$ of $G_1$ were computed for various levels of alteration. Regarding the edge weight**,** a uniform random number ([0, 50]) was added to the initial weight of $G_1$ edges. For node insertion**,** the alteration level was defined as the number of nodes added to $G_1$. Finally, for the spatial location, the alteration level was defined as a random shift of each node of the altered graph $G_2$ to one the possible surrounding positions, either close to (low alteration level) or farther from (high alteration level) the corresponding node in the initial graph $G_1$.

These steps were applied 1000 times for each type of alteration and results were averaged for each method.

**F. Real data**

In order to evaluate the performance of SimiNet on real data, we used brain networks identified during a visual task. Dense electroencephalographic (EEG) data were recorded when subjects named pictures presented on a screen. Pictures were selected from the Snodgrass database (ALARIO AND FERRAND, 1999). Two categories of visual stimuli were shown: tools and animals. The number of pictures for each category was chosen equal (n=37) and several psycholinguistic parameters were controlled to get equivalent datasets (name agreement, image agreement, age of acquisition as well as linguistic parameters like oral frequency, written frequency, letters/phonemes/syllables and morphemes numbers, *see supplementary materials* Fig. S1 and Table S1). The recorded signals were then processed using an EEG source connectivity method recently developed to identify cortical brain networks from scalp EEG data (Hassan et al., 2015a; Hassan et al., 2014; Kabbara et al., 2017). Following this method, brain networks reconstructed from EEG signals (filtered in the gamma frequency band) and corresponding to each category were obtained for twenty one participants (11 women: mean age 28 year; min: 19, max: 40 and 10 men: age 23 years; min: 19, max: 33). For the purpose of this work, we analyzed the networks identified at two different time windows corresponding to two distinct steps of the cognitive process: visual processing (1-119 ms) and access to memory (151-190 ms). These windows were obtained using a clustering algorithm allowing for segmentation of the cognitive process (from picture display to naming) (Mheich et al., 2015). During the first window, (1-119 ms), we expect that the similarity index between brain networks of the two categories of pictures will be high for two reasons, at least. First, it has been shown with fMRI that the primary visual cortex (along with the sensory motor cortex) is certainly the less variable region when functional connectivity is measured between individuals (Mueller et al., 2013). The authors demonstrated that variability of functional connectivity increases from unimodal cortices to multimodal association cortices and correlates positively with the proportion of long range functional links. These results were shown both in their own analysis and in a meta-analysis they ran while grouping 15 studies measuring functional connectivity with fMRI. The second reason why variability should increase with time across different periods of the cognitive process is linked to the interference of semantic judgment on brain processing. During the very first steps of visual processing of picture, semantic has little or nothing to do with the brain operations aiming at reconstructing the image. Moreover, the picture naming task does not require the participants to make a choice from the stimulus (like in a go/no-go task) or to categorize animals or faces which could have led to strong semantic interference around 70 to 80 ms post onset (Vanrullen and Thorpe, 2001). In consequence, this should result in stronger similarity indices for the first period (visual processing) as compared with the second one (access to memory). In addition, we expect a lower similarity index between networks during the second window during which the participants are able to consciously manipulate the concepts shared by the pictures (see the discussion for more detail). The five similarity indexes were calculated between the networks obtained over the two time windows when tools or animals where named by subjects. We evaluated whether SimiNet (compared to other algorithms) was able to discriminate the 2 categories. This study was approved by the Committee for the Protection of Persons (CPP) and a local Ethics committee, (conneXion study, agreement number 2012-A01227-36, promoter: Rennes University Hospital). In this application graph $G$ is defined as a set nodes $V$ representing the brain regions segmented from a Destrieux Atlas (Destrieux et al., 1998) and the edges $E$ represent the functional connectivity between regional time series. In the case of brain networks, the geodesic distance was preferred to the Euclidean distance (used in 2D simulations) as it is more suited to the folded brain surface (presence of gyri and sulci). The "Fast-Marching-Toolbox" (Sethian, 1999) was used to compute the geodesic distance.

# Results

**Simulated data**

Results obtained from the comparison of the different algorithms (GED, VEO, DeltaCon, Lambda-distance, and SimiNet) on simulated graphs are shown in Fig. 3. First, regarding the evolution of the similarity indexes ( $sim_{SimiNet}(G_1,G_2)$, $sim_{GED}(G_1,G_2)$, $sim_{\lambda}(G_1,G_2)$, $sim_{VEO}(G_1,G_2)$ and $sim_{DeltaCon}(G_1,G_2)$ ) with respect to gradually-increasing alterations of the edge weights (Fig. 3A), results confirmed that algorithms GED and VEO do not show any change with regard to alterations of the edge weights (similarity=1 for both methods at all alteration levels). This result was expected as both algorithms are based on quantifying common nodes and edges between graphs and they do not take into account the edge weights to measure the similarity. In contrast, the curves obtained for DeltaCon and Lambda-distance decreased dramatically for increasing level of alteration of the edge weight (DeltaCon:

from 1.0 to 0.27 ± 0.07, Lambda distance: from 1 to 0.05 ± 0.02). This result is explained by the fact that the increase of the edge weight results in i) an increase of the eigenvalues for the adjacency matrix of compared graphs and thus ii) a smaller value of $sim_\lambda(G_1,G_2)$. For DeltaCon algorithm, the increase of edge weight will increase the affinity scores between nodes and then the distance between graphs will increase making $sim_{DeltaCon}(G_1,G_2)$ smaller. Interestingly, and in contrast with the previous methods, the SimiNet algorithm disclosed a gradually decreasing similarity index $sim_{SimiNet}(G_1,G_2)$ for gradual alteration of the edge weight.

In Fig. 3B, we show the evolution of the five similarity indexes with respect to gradual insertion of nodes in the altered graph. Typical examples of the simulated networks are also shown (Fig. 3.B, bottom) for the initial graph $G_1$ and for altered versions $G_2$ at level 10 and 18. Results indicate that the five similarity indexes all decreased with the increase of the number of inserted nodes. DeltaCon and VEO showed fairly similar results with a relatively slow decrease rate (VEO: 0.9705 ± 0.01 at level 4 to 0.91 ± 0.02 at level 20; DeltaCon: 0.95 ± 0.004 at level 4 to 0.9028 ± 0.005 at level 20). For the same alteration levels, SimiNet showed a more pronounced decrease of the similarity index (level 4: 0.947 ±0.006, level 20: = 0.73 ± 0.005). In the case of Lambda distance, $sim_\lambda(G_1,G_2)$ values changed from 0.83 ± 0.06 (level 4) to 0.51 ± 0.06 (level 20). Interestingly, for the four above-described algorithms, a linear decrease of the similarity index was observed. Finally, $sim_{GED}(G_1,G_2)$ showed the most marked decrease (nonlinear in this case, GED: 0.28 ± 0.06 at level 4 to 0.09 ± 0.01 at level 20).

In Fig. 3C, the evolution of the similarity index values with respect to alterations in the node positions is presented. Results confirmed that SimiNet is the only algorithm showing sensitivity to this factor. Indeed, values decreased dramatically with the alteration level (level 1: 0.7825 ± 0.05, level 9: 0.095 ± 0.007). Results indicated that similarity indexes computed from the four other algorithms exhibit different values (DeltaCon: 0.397 ± 0.001; GED: 0.042 ± 0.003; Lambda distance: 0.15 ± 0.012; VEO: 0.02 ± 0.01) that did not change with the level of alteration as expected. Results obtained from the comparison with the graph kernel algorithms (Random Walk kernel, shortest path kernel, Weisfeiler-Lehman, Weisfeiler-Lehman shortest path kernel and SimiNet) are shown in Fig. 4. First, regarding the evolution of the similarity indexes ($sim_{RW\,kernel}(G_1,G_2)$, $sim_{WL}(G_1,G_2)$, $sim_{sp\,kernel}(G_1,G_2)$, $sim_{SimiNet}(G_1,G_2)$ and $sim_{WLspdelta}(G_1,G_2)$) with respect to gradually-increasing alterations of the edge weights (Fig 4-A).

Results confirmed that algorithms WL, spkernel and WLspdelta do not show any change with regard to alterations of the edge weights (similarity=1 for both methods at all alterations levels).

In return, the curve obtained for RWkernel decreased slowly for increasing level of iteration of the edge weight (from 1 to 0.941 ± 0.012).

In Fig.4B we show the evolution of the graph kernels with respect to gradual insertion of nodes in the altered graph. Results indicate that all similarity indexes decreased with the increase of the number of inserted nodes. Spkernel and RWkernel showed fairly similar results with a relatively slow decrease rate (spkernel: 0.96±0.03 at level 4 to 0.86±0.002 at level 20; RWkernel: 0.98±0.001 at level 4 to 0.82±0.001 at level 20).

WL and Wlspdelta showed similar results with high marked decrease (WL: from 0.6218 ±0.001 to 0.282 ±0.001 at level 20; WLspdelta: from 0.627 ±0.001 to 1.98 ±0.001). In Fig. 4C, the evolution of the similarity index values with respect to alterations in the node positions is presented. Results confirmed that similarity indexes computed from the four kernel algorithms exhibit different values (spkernel: 0.917 ± 0.001; WL: 0.991 ± 0.013; WLspdelta: 0.932 ± 0.012; RWkernel: 0.99 ± 0.001) that did not change with the level of alteration as expected.

**Application to real data**

Results obtained from the application of SimiNet to real brain networks are presented in Fig. 5. The similarity scores between networks identified for the object and animal categories in each of the 21 subjects are represented as two [$21 \times 21$] matrices where lines represent the brain networks (i.e. $G_1$ graphs) associated with object visual stimuli and where columns represent networks (i.e. $G_2$ graphs) associated with animal stimuli. As the similarity index $sim_{SimiNet}(G_1, G_2)$ is symmetric, only the values of the upper triangle are displayed.

As depicted, the first matrix (Fig. 5A, left) shows high similarity values during the first period corresponding to visual processing (1-119 ms). In contrast, lower similarity values (Fig. 5A, right) were observed during the second period corresponding to memory access, (150-190 ms). Typical examples of brain networks with high similarity values at the first period and low similarity values at the second period are illustrated in Fig. 5B where the node size represents the "strength value" a network measure defined as the sum of weights of edges connected to this node. For instance, during the visual processing period, a high similarity ($sim_{SimiNet} = 0.51$) was measured be-tween the network obtained when subject 14 was naming animal stimuli and the network obtained when subject 13 was naming object stimuli. In contrast, during the memory access period, a low similarity index ($sim_{SimiNet}$ =0.04) was computed between the network obtained when subject 16 was naming animal stimuli and that obtained when subject 3 was naming object stimuli. Cells in the diagonal of the matrix represent the similarity indices when comparing objects and animals within subjects. Even in this case, $sim_{SimiNet}$ is higher during the first (average of similarity values in the diagonal: 0.1735) period than the second one (average of similarity values in the diagonal: 0.0914). Fig. 5C shows the boxplots of three similarity indexes ($sim_{SimiNet}(G_1,G_2)$, $sim_{DeltaCon}(G_1,G_2)$, $sim_{GED}(G_1,G_2)$) obtained by the three corresponding methods (SimiNet, DeltaCon and GED) during the two periods (visual processing and memory access). The three methods indicate a decrease in the inter-conditions (objects vs. animals) similarity values. During visual processing (1-119 ms), median value of 0.1, 0.22 and 0.14 were observed for SimiNet, DeltaCon and GED, respectively. As expected, these values decreased during the memory access period (151-190 ms) with median values of 0.03, 0.21 and 0.12 for the three methods, respectively. However, and interestingly, SimiNet is the only method showing a statistically significant difference between the two periods (Wilcoxon test, $p<0.01$).

# Discussion

A challenging issue in brain research is to measure the similarity between two spatially-defined networks. In this paper, a new algorithm, called SimiNet, was proposed to calculate the similarity between two graphs with known coordinates for nodes. The proposed algorithm is based on the calculation of i) a distance between nodes in the two graphs, itself based on a least cost sequence of changes (substitution, insertion and deletion of nodes) that are necessary to map one graph on the other and ii) the difference edge weights between the two graphs. The proposed algorithm was compared with eight previously-published graph similarity algorithms (GED, DeltaCon, VEO, $\lambda$-distance, Random Walk kernel, shortest path kernel, Weisfeiler-Lehman, Weisfeiler-Lehman shortest path kernel). SimiNet was shown to improve results in simulated situations involving a shifting of the location of nodes. SimiNet showed also a higher performance in comparing real brain networks obtained from dense EEG during a cognitive task consisting in naming items of two different categories (objects, animals). These findings are discussed hereafter.

*Network similarity under spatial constraint*

Measuring similarity among networks is a topic of increasing interest (Schieber et al., 2017; Shimada et al., 2016; Van Wijk et al., 2010). Several approaches have been proposed to compare networks, in various application domains (social networks, biology, bioinformatics...). The techniques used for comparing brain networks can be classified into three categories: i) *global*, consisting in comparing between global graph measures (degree, modularity, hubs…) computed from the two networks (Stam et al., 2009; Stam and Reijneveld, 2007), ii) *node-wise*, consisting in computing graph metrics for each node of the networks such that multiple comparisons can be assessed (Supekar et al., 2009) and iii) *edge-wise*, consisting in comparing all the available edges in the networks, an approach called Network Based Statistics (NBS) (Zalesky et al., 2010). However, the spatial location of nodes is not accounted for in these approaches. Recently, the spatial constraint was considered in a new metric for computing graph measures in brain networks analysis (Pineda-Pardo et al., 2014). Authors showed that this metric could distinguish the global connectivity of structural networks from functional networks only when the physical locations of nodes are considered. Along the same line, we propose a new algorithm to measure the sim-ilarity between graphs. The key feature of this algorithm is that it takes into account the physical locations of the network nodes. The results of SimiNet applied to real data confirmed the importance of including the physical location of nodes for assessing the (dis)similarity of brain networks involved into two distinct steps of a cognitive task.

Other algorithms (baselines) investigated in this study showed lower performance in detecting the similarity when the two networks do not have the same number of nodes or the spatial location of the nodes is changing. Note that this does not detract the importance of these algorithms as they were developed for specific applications where the spatial location of graphs is a non-relevant factor. Nevertheless, a comparison between modified versions of the baselines (adding to each of the algorithm –if possible- the spatial information) and SimiNet is of interest and could be the subject of further investigation.

*Methodological considerations*

First, the distance used between nodes in the simulation case was assumed to be Euclidian. This distance is not fully appropriate in the real case as brain networks take place over a folded brain surface which consists of sulci and gyri. For this reason, the geodesic distance was preferred to the Euclidian distance. Second, a crucial parameter to be tuned in SimiNet is the radius of the disk used to detect the neighbors of a given node. An increase of the radius $R$ will automatically lead to an increase of the similarity index between the two compared graphs.

In the simulation case, $R$ was chosen to be equal to 1.5 as representing the minimal Euclidian distance between two nodes in the grid (example of Fig. 1). In the real case, $R$ was chosen as the average distance between all nodes. As we were aware about the effect of $R$ on the similarity value, comparative analyses were performed for different values of $R$ (the minimal distance and the maximal distance between two nodes). The results (not shown here) indicated that the performance of SimiNet with respect to the other tested algorithms is preserved, whatever the value of $R$ provided that the condition (Cost (substitution) < Cost (Insertion) + Cost (Deletion)) is preserved.

Third, the current version of the developed algorithm can be applied only between two graphs if their nodes are in the same reference axis frame. Thus, the cost of the three operations: insertion, deletion and substitution of nodes in both graphs will be the same. In the case where the two graphs are not measured in the same axis frame, a transformation is needed as a preprocessing step before applying SimiNet.

From the application viewpoint, brain networks are usually defined in the same spatial reference system. Corresponding graphs are located on a predefined coordinate system, typically 3D mesh of the brain obtained from the subject's structural magnetic resonance imaging (MRI) data. Nevertheless, in some specific applications, brain graphs can be represented in different space. In a recently-published study, we dealt with this issue: in order to apply SimiNet to distinct graphs initially defined in two different coordinate systems (3D vs. 2D), we used a transformation (projection from volumetric to surface coordinates) to bring considered graphs in the same referential (Hassan et al., 2017; Hassan et al., 2016).

Fourth, the main tackled issue in the proposed algorithm is to take into account the physical location of nodes when comparing two graphs. This is a crucial feature in the context where compared graphs correspond to brain networks. Consequently, the proposed algorithm first focuses on the coordinates of nodes (node distance) to match both graphs under comparison. In the second step (edge distance), the algorithm also accounts for the characteristics of edges (their weights), a feature that describes the degree of connectivity (weak to strong) among nodes (i.e. brain areas). Interestingly, the proposed algorithm also considers another feature: the node neighborhood, more specifically, in the first step (node distance), SimiNet makes use of this neighborhood (defined as a disk of radius $R$ centered on the considered node) in order to select between the insertion, suppression or substitution operations. Nevertheless, we believe that taking into account other network measures into consideration to match the two graphs could be of interest such as features dealing with the network segregation (local features such as clustering coefficient, node edge angles(Armiti and Gertz, 2014)) or integration (global features such as the 'hubness' of a node using its degree or its centrality). However, the objective would be different from that dealt in this paper, focused on topology.

Fifth, there is an inter-subject variability in brain anatomy (size and shape for instance). To deal with this variability when comparing brain networks, two solutions can be considered depending on the context of the study (subject-specific or not), in the case where structural MRI data is available for each subject, 3D brain meshes should be normalized before application of SimiNet. In the case when structural MRI is not available for each subject, a template (average brain) can be used for all subjects (which is the case in our study here). In both cas-es, the processed brain networks are represented at the exact same scale and thus brain scaling effects are not encountered.

*Object categorization in the human brain: a network-based approach*

SimiNet was originally developed to analyze the similarity between brain networks involved in cognitive tasks. In this study, the EEG source connectivity analysis allowed us to identify brain networks at cortical level from dense-EEG scalp recordings (Hassan et al., 2015a). Different brain networks were identified for two different categories of stimuli (objects vs. animals). Our intent was to assess the capability of SimiNet and other tested methods to detect significant differences in identified networks.

To our knowledge, this study constitutes the first attempt to assess object categorization in the human brain from a network-based approach using dense EEG source connectivity. During the picture naming task, we detected significant difference between networks identified during the time period associated with 'visual processing' and that related to the 'access to memory'. During visual processing, the networks were mainly occipital involving the inferior occipital, the lateral occipito-temporal sulcus and occipital pole. This period was shown to be related to the visual feature extraction preceding the object category recognition (Thorpe et al., 1996; Vanrullen and Thorpe, 2001). In the present task that just consists in naming pictures overtly, we don't ask the participant to detect animals or to categorize visual scene. As a consequence to this, influences from semantic information on earlier visual processes are low or have not started yet. Moreover, most of the period corresponding to the first time-window precedes the N1, a component of ERP studies that peaks around 150 ms well known to be the first component sensitive to semantic modulations. Obviously some participants can be aware of viewing an animal as soon as 70 to 80 ms after the stimulus onset (Vanrullen and Thorpe, 2001) but it could remain unconscious until its features are mapped onto a memorized concept. These considerations explain the high similarity values observed between object and animal networks during this time period. For the second period, results showed a network involving the occipital regions but with an implication of the bilateral inferior temporal sulcus. This network is known to be related to semantic working memory system when someone tries to remind the name of the presented object (Martin and Chao, 2001). This period was considered as the first instant of categorization in the human brain (Vihla et al., 2006). More cautiously, this can explain the significantly lower similarity values between graphs for animals and objects which could then share different neural substrates. This latter result could also explain the statistical significant difference between the two periods; the similarity index of the second period being lower than the similarity index of the first one.

Two less cognitive but more plausible explanations of these effects could be that i) variability in general increases with time after stimulus onset: indeed, time response latencies, inter-subject differences and attentional level are known to fluctuate (the farther from the stimulus onset, the higher the variability is) and that ii) the second period is shorter in duration than the first period. During the stage of brain network identification, graphs that served to segment the period between 150ms and 190ms are less robust than those which allow identifying the first period. They poorly contribute to the global explained variance (49% vs 63% for the first period) despite remarkable good presence indexes (81% vs. 80 % for the first period).

Apart from which of the previous explanations created these effects (similarity index changes and its significant decrease), both effects were captured by the SimiNet algorithm while the other algorithms failed. As the other algorithms showed no significant difference, we assume that the difference between the networks identified over the two periods is probably related to variations in the spatial location of the nodes, a feature taken into account only by SimiNet.

Moreover, in this paper we evaluated the performance of SimiNet on two different categories: objects and animals. In line with the work of Gallant et al (Huth et al., 2012; Kay et al., 2008), these findings could be extended to build a 'semantic space' of the visual stimuli (human vs non-human, mobile vs immobile, social vs nonsocial …) based on the similarity index calculated using SimiNet. Because this interpretation remains quite speculative in this paper, we are currently addressing the question of early (not semantic) vs late processes (semantic) with a new dataset using different modalities (visual and auditory). This will allow us to get the reverse pattern of similarity index from the first time period with very dissimilar graphs due to different modalities to the second time period with more similar graphs as the semantic system is shared between modalities.

# Conclusion

In this paper, a new algorithm, called SimiNet, for quantifying the similarity between networks under a spatial constraint (position of nodes) was proposed. On simulated graphs, this algorithm showed higher performance than eight state-of-the-art algorithms in detecting some shifts in the node location. When applied to real EEG data, SimiNet could detect significant differences in brain networks associated with two different categories of pictures (objects and animals) used in a cognitive task. We believe that the proposed algorithm can be useful in pattern analysis problems involving a quantification of the similarity between graphs in which the physical location of the nodes is a key parameter.

# Acknowledgement

Tthis work has received a French government support granted to the CominLabs excellence laboratory and managed by the National Research Agency in the "Investing for the Future" program under reference ANR-10-LABX-07-01. It was also financed by the Rennes University Hospital (COREC Project named conneXion, 2012-14). This work was also supported by the European Research Council under the European Union's Seventh Framework Programme (FP7/2007-2013) / ERC grant agreement n° 290901.

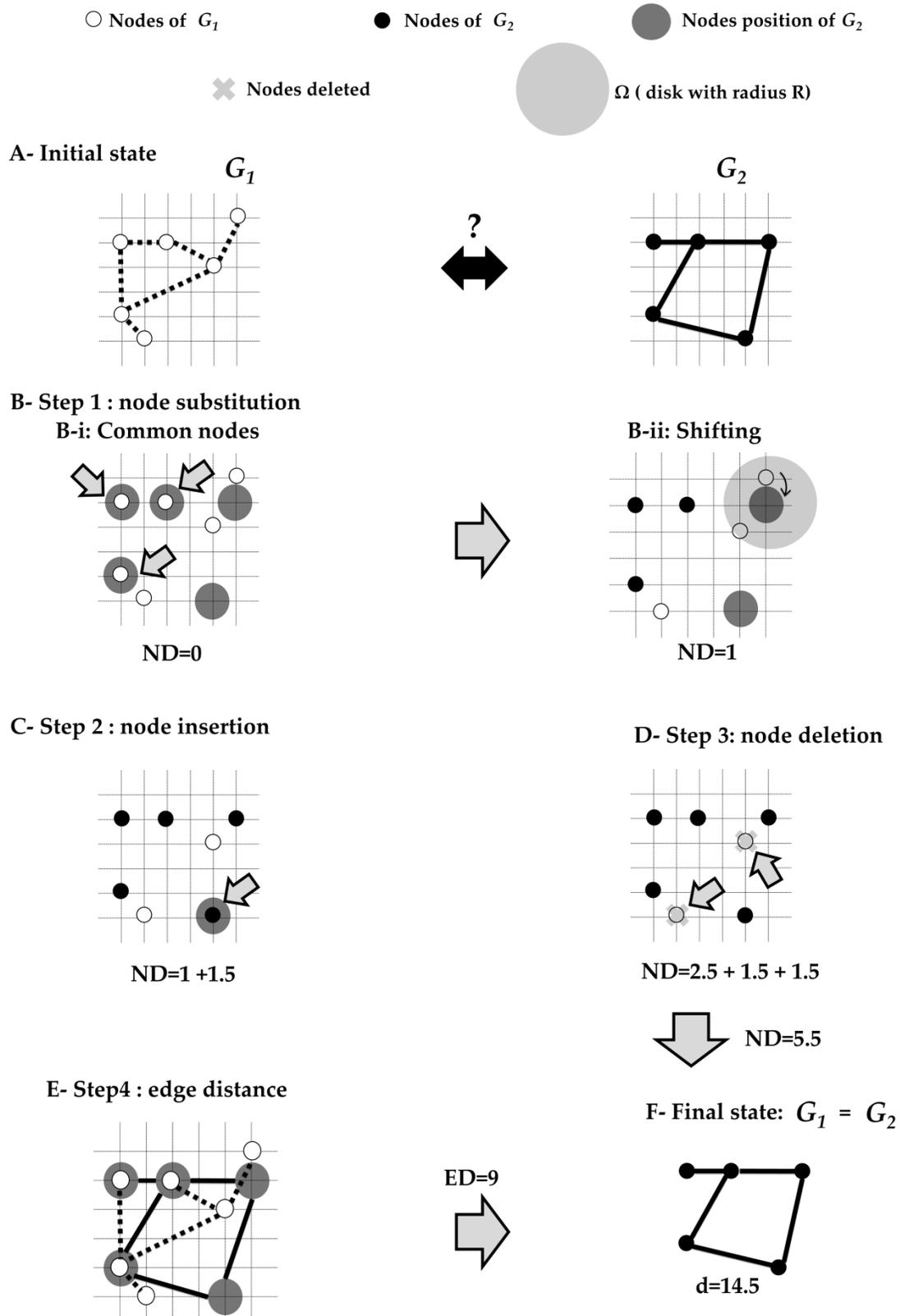

Fig. 1. Illustration of SimiNet algorithm steps. A) Finding the similarity index between $G_1$ and $G_2$. B) Step1-i: detection of nodes common to $G_1$ and $G_2$. B) Step1-ii: shifting the nearest neighbors located in the defined spatial neighborhood (disk with radius $R$ =1.5), shifting cost equals 1. C) Step 2: insertion of new nodes when no neighbor is found within the neighborhood (cost of insertion = $R$ =1.5). D) Step 3: deletion of remaining nodes of $G_1$ (deletion cost = insertion cost =1.5). E) Step 4: computing the edge distance between $G_1$ and $G_2$. F) Final state: $G_1$ matches $G_2$ and $d(G_1,G_2)= ND+ED = 14.5$.

*Algorithm SimiNet*

INPUT: $A_{G_1}$, $A_{G_2}$, $R$
    // Node distance
1:  Initialize $ND = 0$
2:  for each node $v_2 \in V_2$
3:    if $\exists v_1 \in V_1 \,/\, \text{distance}(v_1, v_2) < R$
4:      shifting $v_2 \leftarrow v_1$
5:      $ND = ND + \text{distance}(v_1, v_2)$
6:    else
7:      // insertion a node in $G_1$
8:      $ND = ND + R$
9:    end if
10:  end for each
11:  delete the remained nodes in $G_1$ and update $ND$
    // Edge distance
12:  Initialize $ED = 0$
13:  for $k = 1 \to T-1$
14:    for $p = k+1 \to T$
15:      $ED = ED + \left| W^{G_1}_{v_1^k, v_1^p} - W^{G_2}_{v_2^k, v_2^p} \right|$
16:    end for
17:  end for
18:  $d = ND + ED$
19:  $sim = (1 / (1 + d))$
Return: $sim$

Fig. 2. Pseudo-code of the proposed SimiNet algorithm

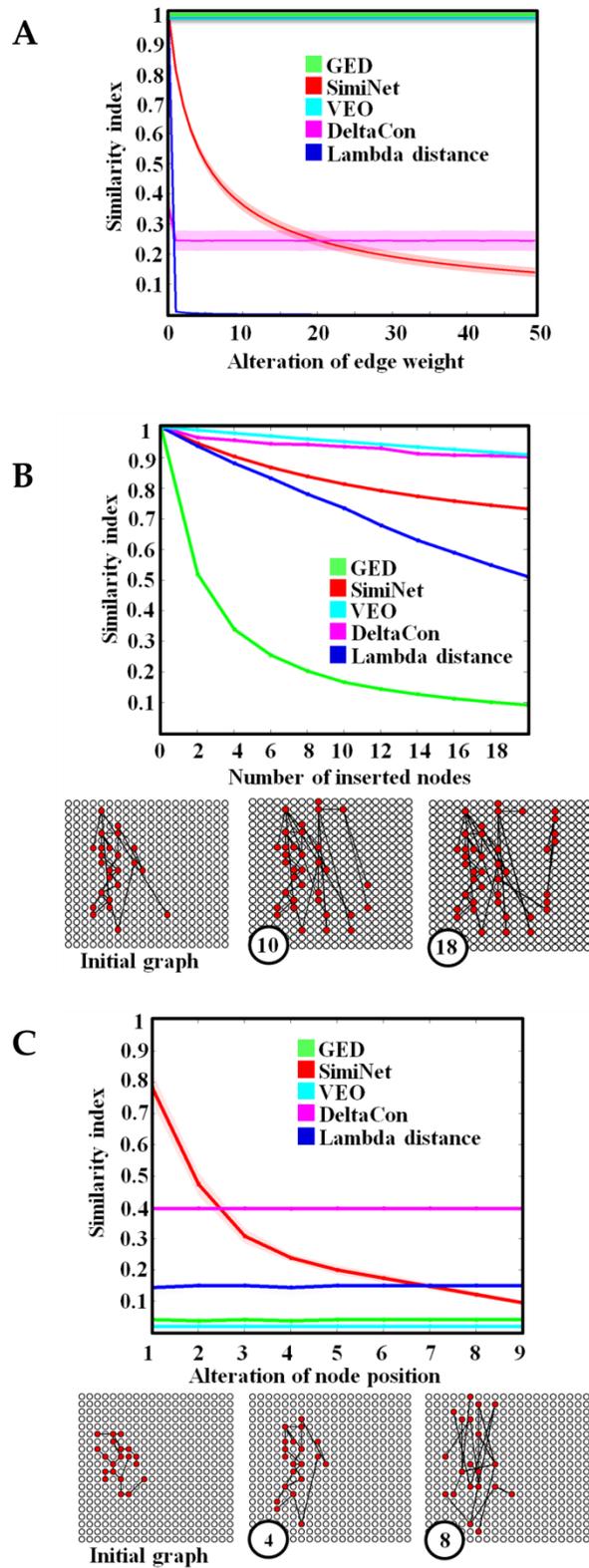

Fig. 3. Variation of similarity indexes computed from the 5 algorithms under evaluation with respect to three types of graph alterations: A) Increase of edge weights. B) Insertion of nodes C) Shifts in the spatial location of nodes. The dark color represents the average value of the similarity measures while the shadowed area represents the standard deviation.

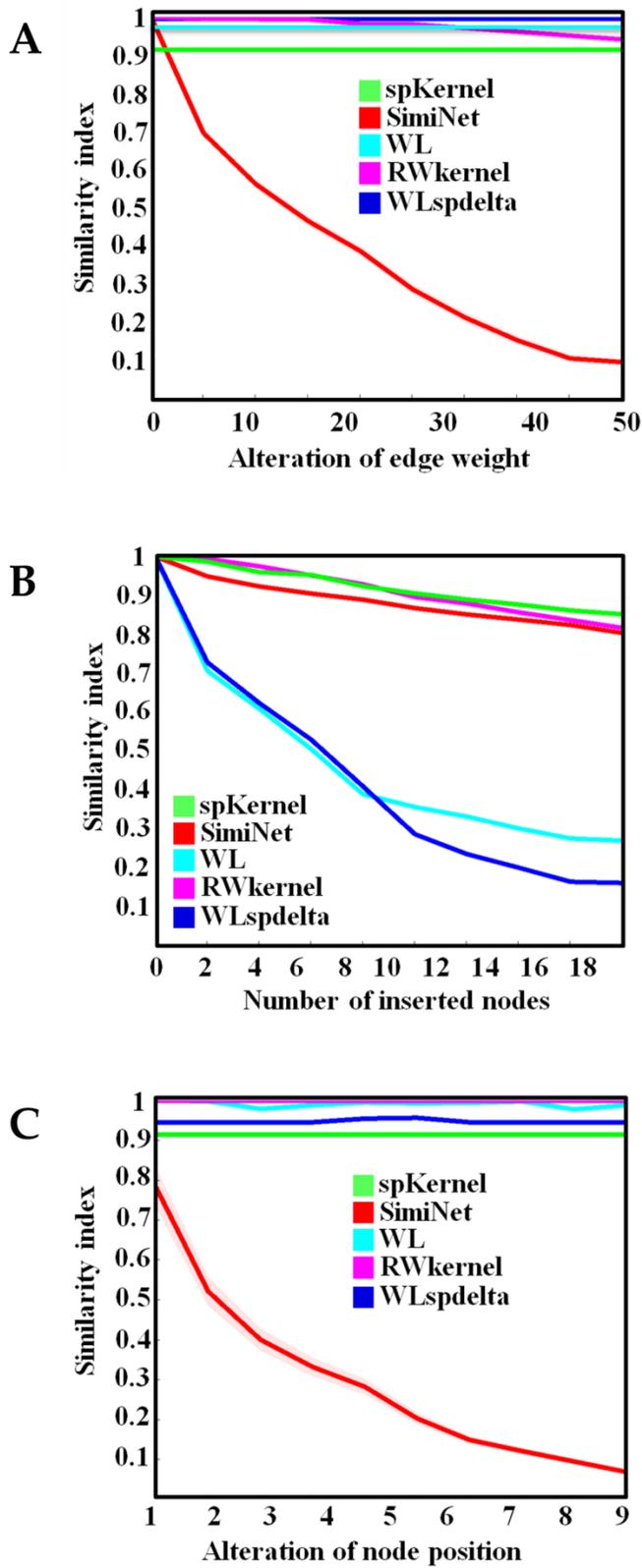

Fig.4: Variation of four graph kernel algorithms under evaluation with respect to three types of graph alterations: A) Increase of edge weights. B) Insertion of nodes C) Shifts in the spatial location of nodes. The dark color represents the average value of the similarity measures while the shadowed area represents the standard deviation

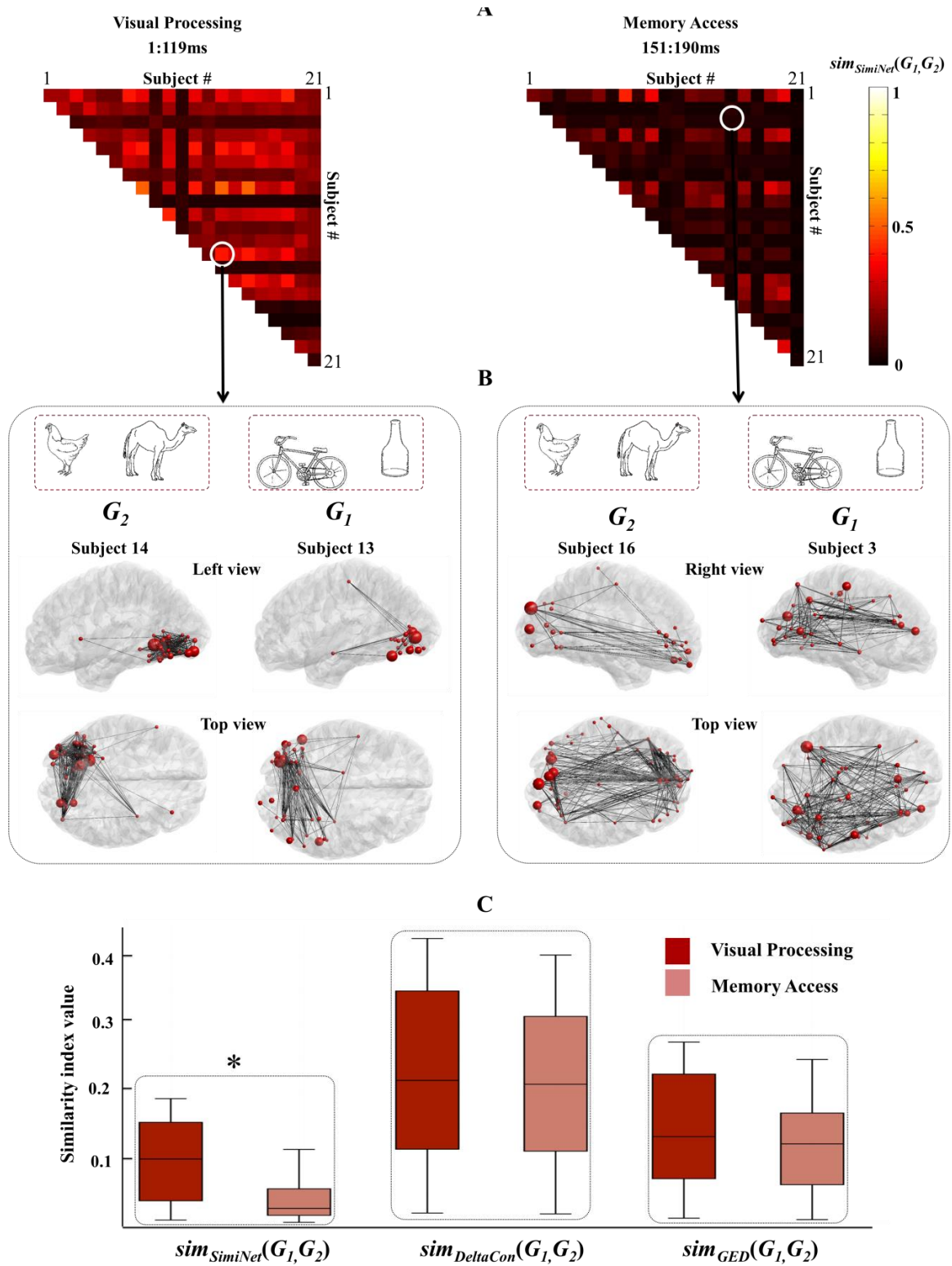

Fig. 5. A- Inter-subject variability of the similarity index ( $sim_{SimiNet}(G_1,G_2)$ ) on real brain networks identified from EEG where $G_1$ and $G_2$ represent respectively the connectivity graphs of the subjects during tools and animals picture naming. Left: value of connectivity graphs in the first period (1-119ms). Right: similarity values of connectivity graphs during the second period (151-190ms). B- The connectivity graphs: 3D representation for 2 different subjects. C- Boxplots show significant difference of similarity values between the two first periods of the cognitive process using SimiNet. Networks were obtained and visualized using EEGNET (Hassan et al., 2015b)
.

TABLE 1
DESCRIPTION OF THE NOTATIONS USED IN THIS PAPER

| Notation | Description |
|---|---|
| $Gr, T$ | Grid, total number of nodes |
| $x_v, y_v$ | Abscissa and ordinate for a node $v$ |
| $\Omega$ | Spatial neighborhood |
| $G$ | Graph |
| $V, n$ | Set of nodes, number of nodes ($n \leq T$) |
| $E, m$ | Set of edges, number of edges |
| $K_v$ | Index of node $v$ |
| $v_i^{K_{v_i}}$ | Node of graph $G_i$ with index $K_{v_i}$ |
| $sim(G_1, G_2)$ | Similarity index between $G_1$ and $G_2$ |
| $d(G_1, G_2)$ | Distance between $G_1$ and $G_2$ |
| $R$ | Radius of $\Omega$ (when defined as a disk) |
| $A_G$ | Adjacency matrix of graph $G$. $A_G[T,T]$ |
| $ED$ | Edge distance |
| $ND$ | Node distance (based on insertion, deletion, substitution costs) |
| $W_{v_i^k, v_i^p}^{G_i}$ | Weight of the edge between the nodes $v_i^k, v_i^p$ in $G_i$ |
| $diff(W_{v_1^p, v_1^k}^{G_1}, W_{v_2^p, v_2^k}^{G_2})$ | Difference of edges weight between nodes $v_1^p, v_1^k$ in $G_1$ and $v_2^p, v_2^k$ in $G_2$ |